\documentclass[aps,prd,reprint,twocolumn,superscriptaddress,longbibliography,nofootinbib,floatfix,showpacs]{revtex4-1}
\usepackage{epsfig}
\usepackage{amsmath,amssymb,amsfonts}
\usepackage{hyperref}
\usepackage{mathrsfs}
\usepackage{bbm}
\usepackage{slashed}
\usepackage{graphicx}
\usepackage{verbatim}

\usepackage{bm}

\usepackage{bbm}

\usepackage{rotating}

\usepackage[usenames]{color}

\definecolor{darkgreen}{rgb}{0.2,0.6,0}

\newcommand{\be}{\begin{equation}}
\newcommand{\ee}{\end{equation}}
\newcommand{\bw}{\begin{widetext}}
\newcommand{\ew}{\end{widetext}}
\newcommand{\bi}{\begin{itemize}}
\newcommand{\ei}{\end{itemize}}
\newcommand{\bea}{\begin{eqnarray}}
\newcommand{\eea}{\end{eqnarray}}
\newcommand{\ud}{\mathrm{d}}

\newcommand{\LCm}{{\scriptscriptstyle -}} 
\newcommand{\LCp}{{\scriptscriptstyle +}}
\newcommand{\LCpm}{{\scriptscriptstyle \pm}}
\newcommand{\LCmp}{{\scriptscriptstyle \mp}}

\newcommand{\LCperp}{{\scriptscriptstyle \perp}}

\usepackage[T1]{fontenc} \usepackage[latin1]{inputenc}

\begin{document}

\title{The trident process in laser pulses}

\author{Victor Dinu} 
\email{dinu@barutu.fizica.unibuc.ro}
\affiliation{Department of Physics, University of Bucharest, P.O.~Box MG-11, M\u agurele 077125, Romania}

\author{Greger Torgrimsson}
\email{greger.torgrimsson@uni-jena.de}
\email{g.torgrimsson@hzdr.de}
\affiliation{Theoretisch-Physikalisches Institut, Abbe Center of Photonics,
Friedrich-Schiller-Universit\"at Jena, Max-Wien-Platz 1, D-07743 Jena, Germany}
\affiliation{Helmholtz Institute Jena, Fr\"obelstieg 3, D-07743 Jena, Germany}
\affiliation{Helmholtz-Zentrum Dresden-Rossendorf, Bautzner Landstra{\ss}e 400, 01328 Dresden, Germany}

\begin{abstract}
We study the trident process in laser pulses. We provide exact numerical results for all contributions, including the difficult exchange term. We show that all terms are in general important for a short pulse. For a long pulse we identify a term that gives the dominant contribution even if the intensity is only moderately high, $a_0\gtrsim1$, which is an experimentally important regime where the standard locally-constant-field (LCF) approximation cannot be used. 
We show that the spectrum has a richer structure at $a_0\sim1$, compared to the LCF regime $a_0\gg1$. We study the convergence to LCF as $a_0$ increases and how this convergence depends on the momentum of the initial electron. We also identify the terms that dominate at high energy. 
\end{abstract}
\maketitle

\section{Introduction}

The trident process, $e^\LCm\to2e^\LCm+e^\LCp$, in laser fields has recently attracted renewed interest~\cite{Dinu:2017uoj,King:2018ibi,Mackenroth:2018smh,Acosta:2019bvh,Krajewska15,Hu:2014ooa} following~\cite{King:2013osa,Ilderton:2010wr,Hu:2010ye} and the classic experiment at SLAC~\cite{Bamber:1999zt}. While this process was studied already in the early seventies~\cite{Ritus:1972nf,Baier}, it is only with modern high-intensity lasers that it is becoming experimentally important. Although the SLAC experiment is already two decades old and there today exist much stronger lasers, it is still the only experiment directly relevant for this process\footnote{However, trident in a crystal has been experimentally studied in~\cite{Esberg:2010zz}.}. This, however, will likely change soon, given the great deal of interest in this process in itself~\cite{desyWorkshop,LUXEparameters,Abramowicz:2019gvx,MeurenPresentation} and as a first step towards production of many particles in cascades~\cite{Bell:2008zzb,Elkina:2010up,Nerush:2010fe}.    
A related second-order process is double nonlinear Compton scattering~\cite{Morozov:1975uah,Lotstedt:2009zz,Loetstedt:2009zz,Seipt:2012tn,Mackenroth:2012rb,King:2014wfa,Dinu:2018efz,Wistisen:2019pwo}, where the electron instead produces two photons.

We use units with $c=\hbar=m_e=1$, where $m_e$ is the electron mass, and we absorb a factor the charge $e$ into the background field, so $e$ only appears explicitly via the coupling $\alpha$ to the quantized field.    
The trident process depends on $a_0=E/\omega$, where $E$ is the field strength and $\omega$ the characteristic frequency of the plane-wave background field, and $\chi=a_0b_0$, where $b_0=kp$ is the product of the wave vector of the plane wave, $k_\mu$ with $k_0=\omega$, and the momentum of the initial electron $p_\mu$.

The trident process can be separated into one-step and two-step parts, where the latter is obtained from an incoherent product of nonlinear Compton scattering and nonlinear Breit-Wheeler pair production. If the intensity of the background is sufficiently high then one can approximate the trident probability by the two-step where the field is treated as locally constant at the two steps. This locally constant field (LCF) approximation greatly simplifies the calculation and also makes it possible to approximate complicated higher-order processes by a sequence of first-order processes using particle-in-cell codes~\cite{RidgersCode,Gonoskov:2014mda,Osiris,Smilei}. 
However, the LCF approximation can break down at both low and high energies~\cite{Dinu:2015aci,DiPiazza:2017raw,Podszus:2018hnz,Ilderton:2019kqp}. 
Moreover, this also means that one misses much of the additional structure in the trident probability that is not already in the LCF approximation of the two first-order processes. This gives motivation for studying fields with moderately high intensity. 
 
In this paper we consider the trident process at moderately high intensity, i.e. $a_0\gtrsim1$. In this regime we can neither make a perturbative expansion in $a_0$ nor employ the LCF approximation, which is basically an expansion in $1/a_0\ll1$~\cite{Dinu:2017uoj}. 
As we showed in~\cite{Dinu:2017uoj}, for $a_0\sim1$ the one-step terms are in general on the same order of magnitude as the two-step term, and then one in general has to include the exchange term, i.e. the cross-term between the two terms in the amplitude that are related by exchanging the two electrons in the final state. The exchange term is much harder to calculate than the direct terms, i.e. the non-exchange terms\footnote{Note ``direct'' does not mean ``instantaneous'' or ``one-step''. We define ``direct'' as the complement of the exchange term, and it gives one part of the one-step but also the two-step.}. 
So, it is important to know in which parameter regimes one can neglect the exchange term. We study this here. 

While the trident process\footnote{Note that~\cite{Bamber:1999zt} used different notation for the different contributions to this process. Here trident includes the incoherent product of nonlinear Compton and Breit-Wheeler as well as everything else.} has been observed in~\cite{Bamber:1999zt}, there the laser only had a small $a_0$ and they found a probability that scaled as $a_0$ to the power of the number of photons that need to be absorbed to produce a pair. While that observation was already nonlinear in $a_0$, for higher intensities the dependence on the field strength becomes more interesting. 
For $a_0\gg1$ and $\chi\ll1$ one finds a probability that scales as~\cite{Ritus:1972nf,Baier}
\be\label{compSchwinger1}
\mathbb{P}\sim\exp\left(-\frac{16}{3\chi}\right) \;.
\ee  
This scaling is basically the same as for the nonperturbative Sauter-Schwinger process~\cite{Sauter:1931zz,Heisenberg:1935qt,Schwinger:1951nm}, cf.~\cite{Bamber:1999zt}, where the probability of spontaneous pair production by a constant electric field scales as
\be\label{compSchwinger2}
\mathbb{P}\sim\exp\left(-\frac{\pi}{E}\right) \;.
\ee 
Since in the trident case we have $\chi=E$ in the rest frame of the initial electron, the difference between~\eqref{compSchwinger1} and~\eqref{compSchwinger2} is basically just a numerical factor. 
There is a similar scaling for nonlinear Breit-Wheeler pair production~\cite{Reiss62,Nikishov:1964zza,Hartin:2018sha}
\be
\mathbb{P}\sim\exp\left(-\frac{8}{3\chi_\gamma}\right) \;,
\ee
where $\chi_\gamma=a_0kl$ is proportional to the momentum $l_\mu$ of the initial photon. However, in this case there is no frame in which $\chi_\gamma$ is equal to the field strength, it always contains the frequency of the initial photon. The advantage of Breit-Wheeler, as well as other photon/field-assisted mechanisms~\cite{Schutzhold:2008pz,Dunne:2009gi}, is that one has production of matter from an initial state with only photons. The advantage of the trident process is that one finds an exponential scaling with an even closer resemblance to the elusive Sauter-Schwinger process. This provides further motivation for studying the trident process with upcoming high-intensity laser facilities, such as LUXE~\cite{LUXEparameters,Abramowicz:2019gvx} and FACET-II~\cite{MeurenPresentation}.   

This can be generalized to inhomogeneous fields. In the trident case we find~\cite{Dinu:2017uoj} for $a_0\gtrsim1$ and $\chi\ll1$
$\mathbb{P}\sim\exp\left(-f(a_0)/\chi\right)$ with some function $f$ which depends on the field shape, 
and in the Sauter-Schwinger case one finds, for a time-dependent electric field with frequency $\omega$ and $E\ll1$,
$\mathbb{P}\sim\exp\left(-F(\gamma)/E\right)$, 
where the dependence on $\omega$ is in this context usually expressed in terms of the Keldysh parameter $\gamma=\omega/E$, which corresponds to $1/a_0$.

In our previous paper~\cite{Dinu:2017uoj} we studied the trident process in plane-wave background fields with the lightfront approach and by integrating over the transverse momentum components as well as summing over all the spins. The advantage of this approach is that the result only depends on relatively few parameters, viz. the field parameters and the longitudinal momentum components. 
This approach also allowed us to derive relatively compact expressions for the probability for arbitrary field shapes. In this paper we will use these expressions as a starting point for further investigation into the relative importance of the various terms.
  
This paper is organized as follows. In Sec.~\ref{DefinitionSection} we give some basic definitions.  
In Sec.~\ref{DISC} we study numerically the trident spectrum for a pulsed oscillating field.
We study the dependence on pulse length, energy parameter $b_0$ and $a_0$.
We have studied the contribution from all one-step terms, including the difficult exchange term, for $a_0\gtrsim1$ i.e. beyond the LCF regime.
In Sec.~\ref{Numerical methods intro} and Appendix~\ref{Numerical methods} we describe our numerical methods.

\section{Formalism}\label{DefinitionSection}

We use lightfront variables defined by $v^\LCpm=2 v_\LCmp=v^0\pm v^3$, $v_\LCperp=\{v_1,v_2\}$ and $\phi=kx=\omega x^\LCp$. The field is given by $a_\LCperp(\phi)$.
The initial electron has momentum $p^\mu$, and the final state electrons and positron have momenta $p_{1,2}^\mu$ and $p_3^\mu$, respectively. The quantities we are interested in here only depend on the longitudinal momentum components of the fermions, which we denote as $b_0=kp$ and $s_i=kp_i/kp$. We also use $q_i:=1-s_i$, $i=1,2$ for the longitudinal momentum of the intermediate photon.  

We consider either the integrated probability $\mathbb{P}$ or the longitudinal momentum spectrum $\mathbb{P}(s)$, which are related via
\be
\mathbb{P}=\int_0^1\ud s_1\ud s_2\theta(s_3)\mathbb{P}(s) \;.
\ee   
These are obtained by performing the Gaussian integrals over the transverse momenta $p_{1\LCperp}$ and $p_{2\LCperp}$. We can perform these integrals analytically for any field shape, so it is a natural step in order to reduce the number of independent parameters. 
From an experimental point of view, it is also worth noting that in an electron-laser collision, the initial electron will in most cases have a very large $\gamma$ factor. This means that the produced particles will travel almost parallel to the initial momentum. So, although we have integrated over all possible values of the transverse momenta, the dominant contribution comes from a relatively small region around the forward direction, see Appendix~\ref{Formulas}. So, performing these transverse integrals effectively gives what a detector around the forward Gaussian peak would measure. So, for high energies the ratios of longitudinal momenta $s_i$ also give us the fraction of the initial electron's momentum/energy given to the produced particles.         

The amplitude has two terms, $M_{12}$ and $M_{21}$, where $M_{21}$ is obtained from $-M_{12}$ by swapping place of the two identical electrons in the final state. On the probability level $|M_{12}|^2$ and $|M_{21}|^2$ give what we call the direct part $\mathbb{P}_{\rm dir}$, and $2\text{Re }M_{21}^*M_{12}$ gives the exchange part $\mathbb{P}_{\rm ex}$. Note that ``direct'' does not mean instantaneous. In fact, $\mathbb{P}_{\rm dir}=\mathbb{P}_{\rm two}+\mathbb{P}_{\rm one}^{\rm dir}$ gives the entire two-step $\mathbb{P}_{\rm two}$ as well as one part of the one-step $\mathbb{P}_{\rm one}^{\rm dir}$. The exchange term only contributes to the one-step $\mathbb{P}_{\rm ex}=\mathbb{P}_{\rm one}^{\rm ex}$. These three terms can be compared with the results from other approaches. In our approach we find it convenient to further split $\mathbb{P}_{\rm one}^{\rm dir}$ and $\mathbb{P}_{\rm one}^{\rm ex}$ each into three parts, 
\be
\mathbb{P}_{\rm dir}^{\rm one}=\mathbb{P}_{\rm dir}^{11}+\mathbb{P}_{\rm dir}^{12}+\mathbb{P}_{\rm dir}^{22\to1}
\ee
and
\be
\mathbb{P}_{\rm ex}=\mathbb{P}_{\rm ex}^{11}+\mathbb{P}_{\rm ex}^{12}+\mathbb{P}_{\rm ex}^{22} \;.
\ee      
All the contributions to the trident probability are given by the expressions in Appendix~\ref{Formulas}.

\begin{figure*}
\makebox[\textwidth][c]{
\hspace{-.2cm}
\includegraphics[width=1.1\linewidth]{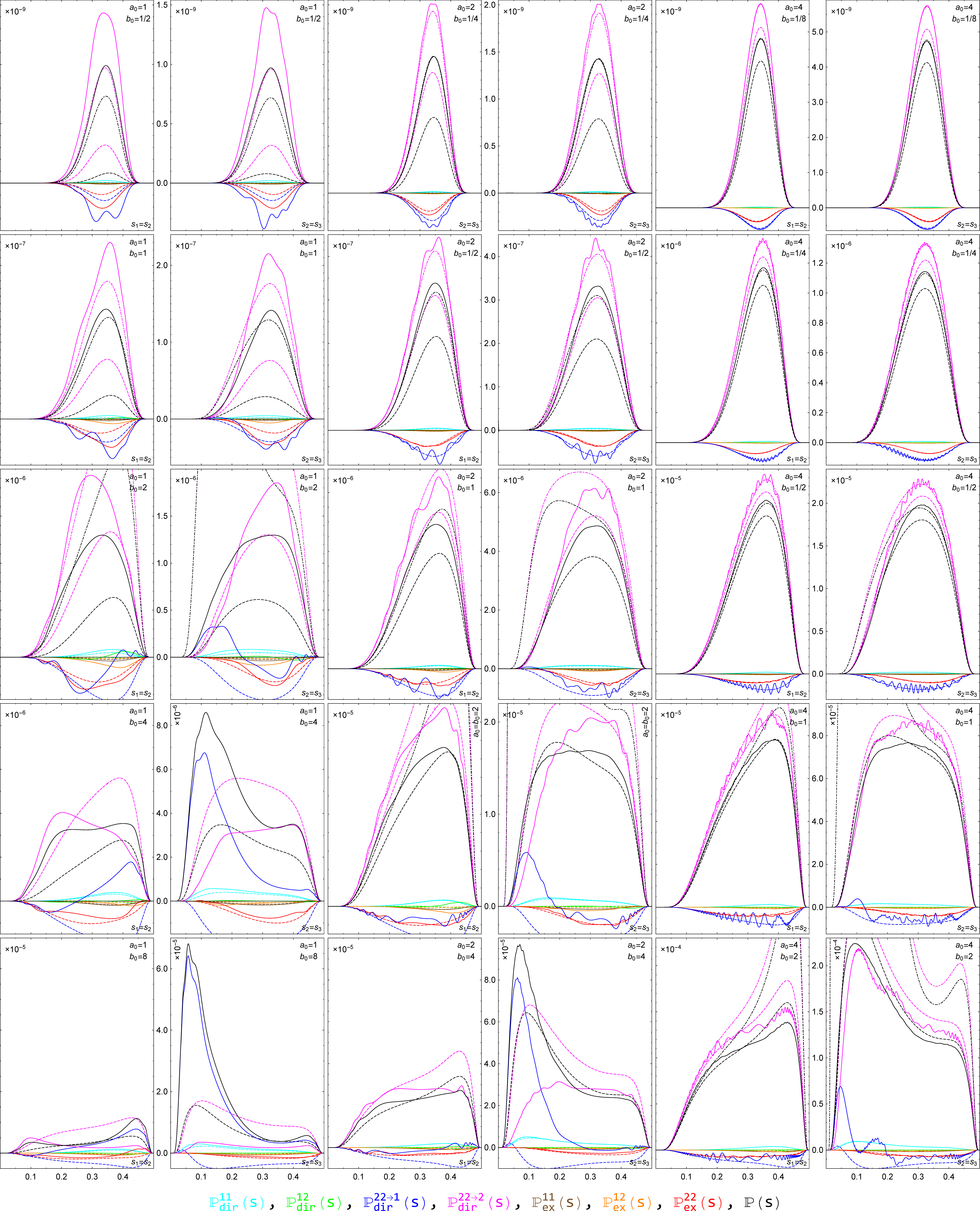}
}
\caption{All contributions for a short pulse, $\mathcal{T}=\pi$. Dashed lines for LCF and dot-dashed lines for LCF+1.}
\label{bothSectionsFig}
\end{figure*}

\begin{figure*}
\makebox[\textwidth][c]{
\hspace{-.2cm}
\includegraphics[width=1.1\linewidth]{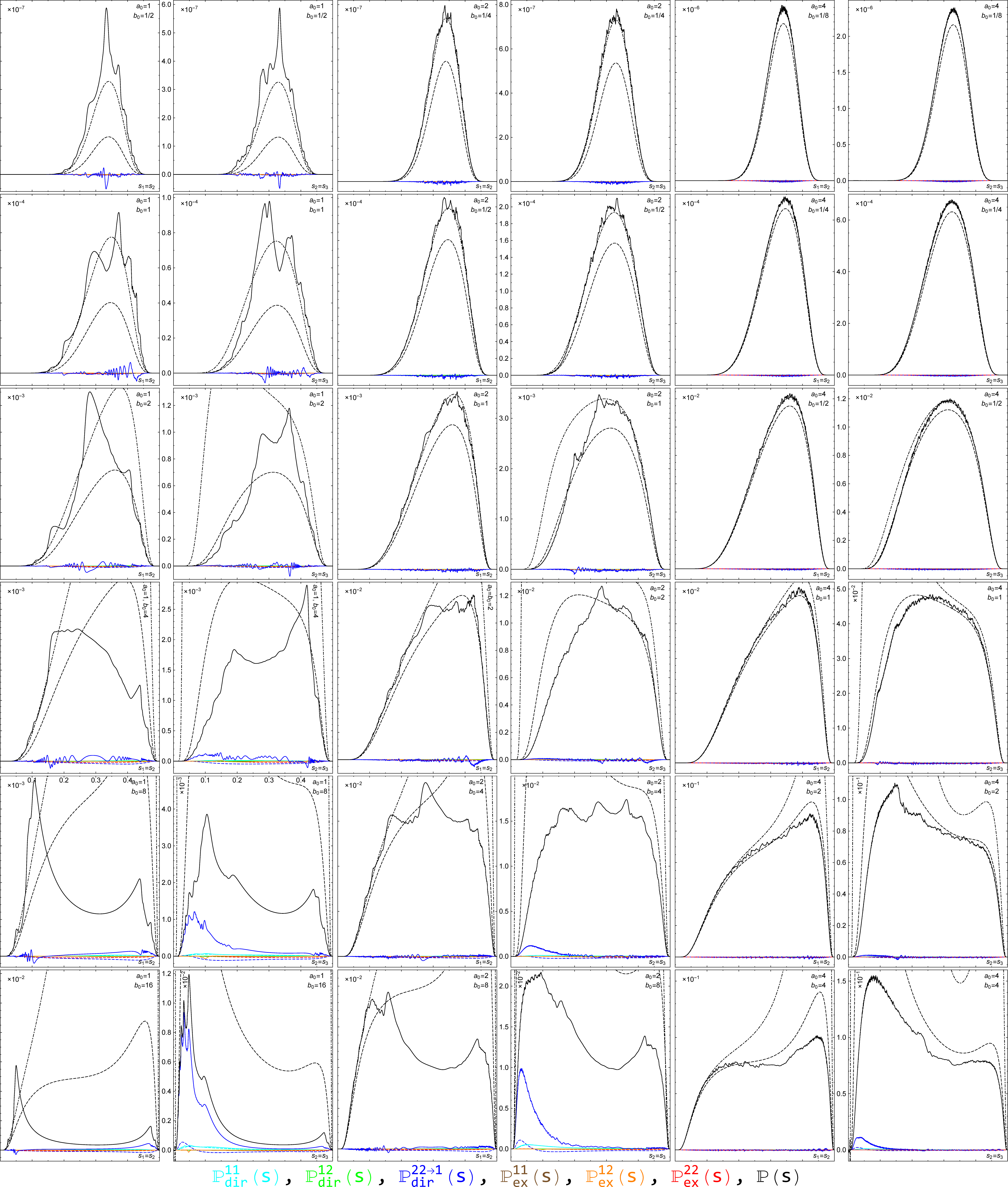}
}
\caption{All one-step contributions and the sum of all contributions for a long pulse, $\mathcal{T}=80$.}
\label{LongPulsePlot}
\end{figure*}

\begin{figure*}
\includegraphics[width=.65\linewidth]{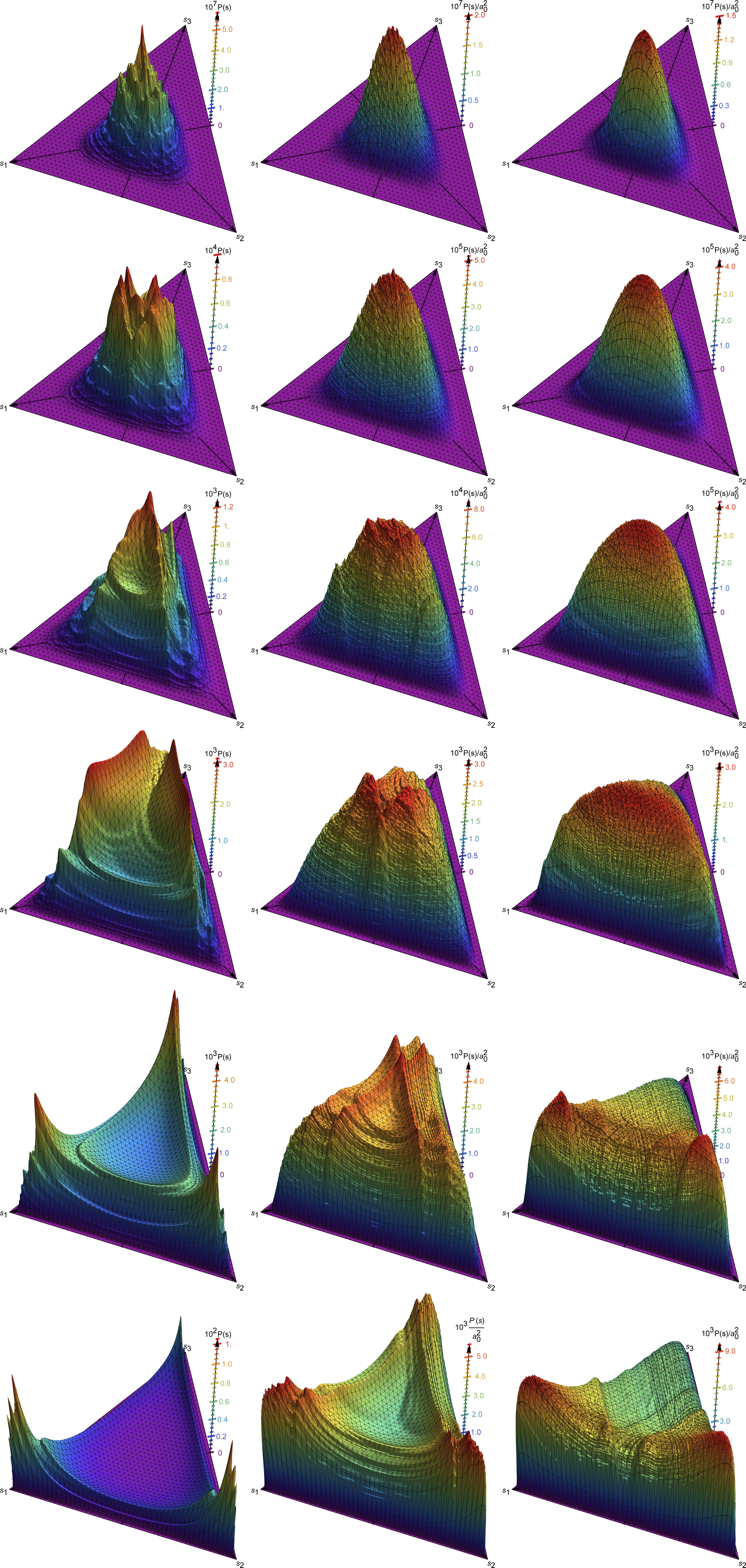}

\caption{The spectrum for the long pulse. $a_0=\{1,2,4\}$ in column $\{1,2,3\}$, $\chi=\{1/2,1,2,4,8,16\}$ from top to bottom.}
\label{3Dlin}
\end{figure*}

\section{Numerical study of the trident spectra}\label{DISC}
In this section we will present a numerical investigation of the trident process for oscillating fields with Gaussian envelope,
\be\label{fieldDef}
{\bm a}(\phi)=a_0\{\cos\xi\sin(\phi-\phi_{0}),\sin\xi\cos(\phi-\phi_{0}),0\}e^{-(\phi/\mathcal{T})^2} \;.
\ee
We may vary the polarization, between linear ($\xi=0$) and circular ($\xi=\pi/4$), as well as the CEP,  $\phi_{0}$. We focus on the choices $\xi=0$ and $\phi_{0}=0$ for they give more structure, provide a tougher numerical challenge and strengthen the importance of one-step terms.
To reveal the importance of pulse length we have considered both an
ultra-short pulse ($\mathcal{T}=\pi$), which maximizes the contribution of the one-step terms, and a longer, realistic pulse ($\mathcal{T}=80$), motivated by state-of-the-art lasers like the ones to be used in LUXE~\cite{LUXEparameters} and FACET-II~\cite{MeurenPresentation}. It might be useful to consider a somewhat shorter pulse if, as in the SLAC experiment, the electron and laser beams collide at a nonzero angle, which would give a shorter effective pulse length. For $a_0\ge1$ and $\chi<1$ we also notice that, unless a pulse with a flat-top envelope is used, the exponential suppression of the rates associated with lower local $\chi(\phi)=a'(\phi) b_0$ values means that only those peaks that are close to the global maximum contribute significantly, hence the field has a shorter effective length.

\begin{figure*}
\includegraphics[width=0.8\linewidth]{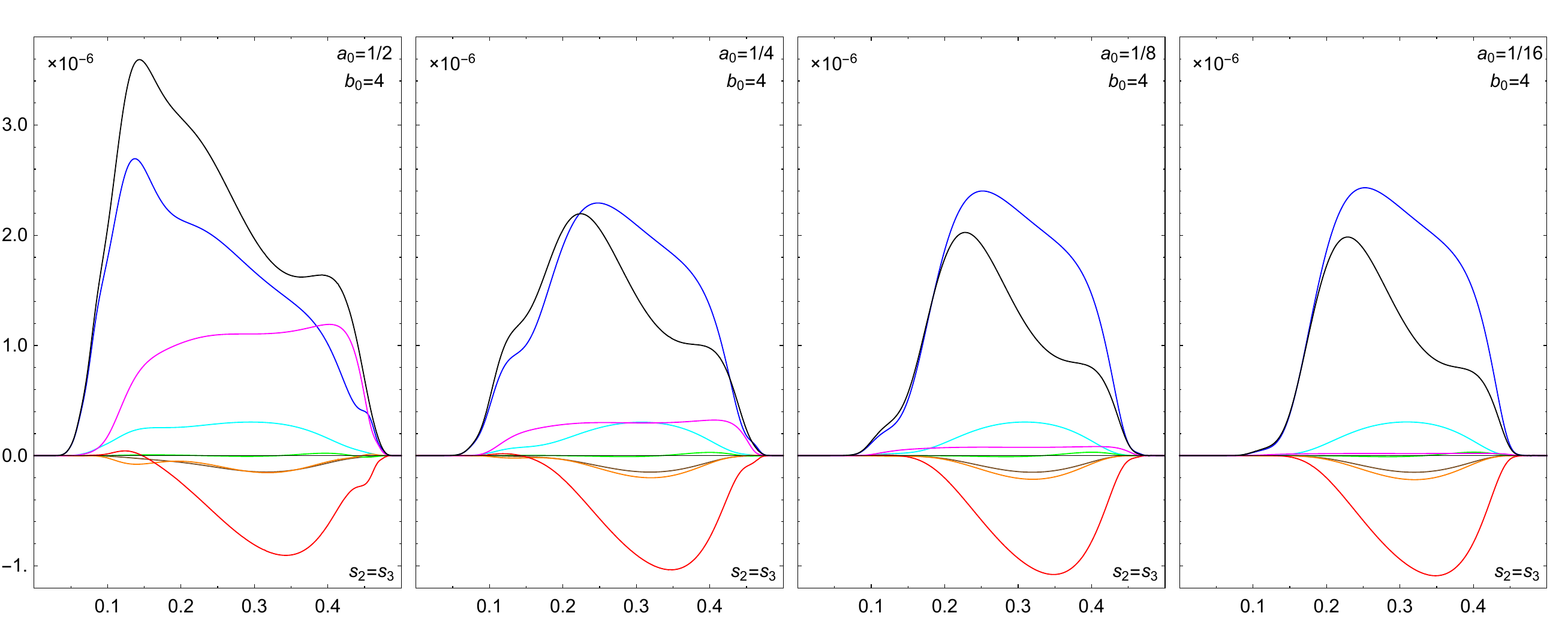} \\
\includegraphics[width=0.65\linewidth]{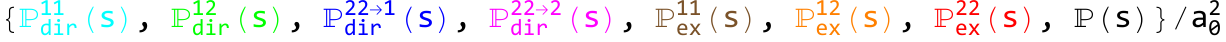}
\caption{Section of the spectra for $b_0=4$ and $\mathcal{T}=\pi$.}
\label{perturbativeplot}
\end{figure*}

Fig.~\ref{bothSectionsFig} shows the $s_{1}=s_{2}$ and $s_{2}=s_{3}$ cross-sections through
the probability density $P\left(  s\right)  $ for the short pulse, detailing
all terms that contribute, their LCF approximation and the improved LCF+1
approximation. LCF+1 is obtained by including the next-to-leading order correction in an expansion in $1/a_0$, cf.~\cite{Dinu:2017uoj,Ilderton:2018nws}.
The values $a_{0}=\left(  1,~2,~4\right)  $ and $\chi=a_{0}%
b_{0}=\left(  0.5,~1,~2,~4,~8\right)  $ are considered.
Fig.~\ref{LongPulsePlot}] shows the same for the long pulse, apart from the two-step term which can be close to
the total probability density and was removed to reduce clutter. Only the one-step terms are shown
explicitely. Fig.~\ref{3Dlin} shows 3D plots of the spectrum for
the long pulse, scaled by dividing by $a_{0}^{2}$.  For Fig.~\ref{LongPulsePlot} and~\ref{3Dlin} we go up to $\chi=16$, as a larger $b_{0}$ value
is needed to find an important contribution of the one-step terms for a longer
pulse. In Fig.~\ref{perturbativeplot} we considered subunitary $a_{0}$ values for the short pulse, scaled by dividing by $a_{0}^{2}$.

To produce the plots in Fig.~\ref{LongPulsePlot}] we approximated the computationally
intensive but small $\mathbb{P}_{\rm ex}^{22}\left(  s\right)  $ by its LCF approximation.
The short pulse results of Fig.~\ref{bothSectionsFig} and the tests we made for longer pulses
suggest that this leads to a small error for the $a_{0}=1$ plots and a very
small one for higher $a_{0}$.  Compare this to the larger discrepancy caused by, for instance, neglecting the focusing of the laser or by the imperfect control of the pulse shape allowed by experiments. Also notice that $\mathbb{P}_{\rm ex}^{22}\left(  s\right)  $ does not posess those high amplitude oscillations which make $\mathbb{P}_{\rm dir}^{22\rightarrow1}\left(  s\right)  $ larger in
magnitude than its LCF values for a long pulse, even at low $b_{0}$. We feel that the modest loss of precision of the result due to the approximation we made hardly justifies the use of the great computing power needed to precisely plot $\mathbb{P}_{\rm ex}^{22}\left(  s\right)$ on the detailed grid, with many $s$ points. 
Considering a long, linearly polarized, pulse with subunitary $a_{0}$ is a computationally
intensive challenge for the future, as $\mathbb{P}_{\rm ex}^{22}\left(  s\right)$ can be very
important in that case and has to be computed accurately for a good precision
of the total result.  The contribution to the spectrum of all the terms but $\mathbb{P}_{\rm ex}^{22}\left(
s\right)  $ can be computed precisely at arbitrarily high resolution, through
the two/three-tiered approach ${\rm A}_2$, building intermediate tables to be reused
again and again. Thus we were able to produce the high resolution 3D plots of Fig.~\ref{3Dlin} for
the long pulse.

We next present the main conclusions we have drawn from our numerical investigations.
The ``glued-up'' term $\mathbb{P}_{\rm dir}^{22}\left(
s\right)  $ is the most important, clearly dominates for a long pulse unless
$a_{0}$ is subunitary. Our interest in the gluing techniques comes from the
conjecture that this can generalize to other processes, see~\cite{Dinu:2018efz,gluingPaper}, including the ones
involving the emission of hard photons.

While good in the LCF limit, the division into steps $\mathbb{P}_{\rm dir}^{22}\left(
s\right)  =\mathbb{P}_{\rm dir}^{22\to1}\left(  s\right)+\mathbb{P}_{\rm dir}^{22\to
2}\left(  s\right)  $ adds spurious oscillations to $\mathbb{P}_{\rm dir}^{22\to1}\left(  s\right)$ and $\mathbb{P}_{\rm dir}^{22\to2}\left(  s\right)$ outside LCF.
They can be seen in the dependence on both \thinspace$a_{0}$ and $s$.
Integrating over $s$ smooths out the former. The same oscillations were seen
for the corresponding ``rates'' defined in~\cite{Dinu:2017uoj}. Hence, as $a_{0}$
increases we see much faster convergence for the undivided $\mathbb{P}_{\rm dir}%
^{22}\left(  s\right)  $ than for $\mathbb{P}_{\rm dir}^{22\to1}\left(  s\right)$ and
$\mathbb{P}_{\rm dir}^{22\to2}\left(  s\right)$ or the ``rate''. Thus, away from
the LCF limit it is natural to study $\mathbb{P}_{\rm dir}^{22}\left(  s\right)  $ as whole. 

However, there is a way to divide integrals discussed in
Appendix~\ref{Numerical methods}, $I=$ $I^{aa}$ $+$ $I^{a0}+I^{0a}$, which is
elegant, efficient and smooth, adding no such oscillations to its terms. It originates in the need to obtain formulas that are regular without the infinitesimal complex shift implemented by $i\epsilon$.  We prefer to use a
more computationally friendly alternative, which only keeps one exponential in the integrals.  The more physically meaningful way to do it involves subtracting field-free counterparts from the factors in the integrand coming
from each step, allowing us to interpret the three terms as describing an
interaction at both steps or just one of them.

The frequency of the oscillations we mentioned increases with $a_{0}$, while
their amplitude slowly decreases. Unlike for the short pulse, for the long one
they are frequent and very irregular, which can be linked to the many sharp
peaks of the derivative introduced by the change of variable $\theta
_{ij}\to\Theta_{ij}$ or to the large number of complex saddle points
with important contribution. All this structure exists for linear
polarization, while for circular polarization (not illustrated), the curves
are much smoother and the computation times needed are significantly shorter, as $\Theta_{ij}$ is a much smoother function.
The rest of the one-step terms converge more quickly to LCF than
$\mathbb{P}_{\rm dir}^{22\to1}\left(  s\right)  $, for linear polarization, as
they lack artificially introduced oscillations. Looking at Fig.~\ref{LongPulsePlot}
we see how much smoother the oscillation structure is for $\mathbb{P}\left(  s\right)
$ than for $\mathbb{P}_{\rm dir}^{22\to1}\left(  s\right)  $. The latter's oscillations are so frequent as to even become
hard to render within the resolution of the plot for larger $a_{0}$. We chose
linear polarization and a symmetric pulse in order to have a worst case
scenario for the consequence of the $\mathbb{P}_{\rm dir}^{22\to1}\left(  s\right)
$ term, as the symmetry makes $\mathbb{P}_{\rm dir}^{22\to2}\left(  s\right)  $
factorize completely into half the product of independent
Compton/Breit-Wheeler probability densities, thus conserving intact and combining all their
oscillations. This maximizes the importance of
$\mathbb{P}_{\rm dir}^{22\to1}\left(  s\right) $ compared to $\mathbb{P}_{\rm dir}^{22}\left(  s\right)$, whose oscillations are partially washed out by the
different choice of step functions that sever the $\Theta_{ij}$ integrals.  Check~\cite{Dinu:2018efz}  for the oscillations shown by the Compton spectra for linear polarization. 

In the expansion in powers of $a_{0}^{-1}$, whose dominant terms are the LCF
results (order $a_{0}^{-2}$ for $\mathbb{P}_{2}\left(  s\right)  $ and $a_{0}^{-1}$ for
$\mathbb{P}_{1}\left(  s\right)  $, dashed lines in the plot), there is also an
$O\left(  1\right)  $ term coming from $\mathbb{P}_{2}\left(  s\right)  $. Including
the latter we obtained an approximation called LCF+1, cf.~\cite{Dinu:2017uoj,Ilderton:2018nws}, which we plotted using
dash-dotted lines. As seen from Fig.~\ref{bothSectionsFig} and~\ref{LongPulsePlot}, the LCF+1 approximation is
generally larger than the LCF one and, at small $b_{0}$, it provides
an obvious improvement. As $b_{0}$ increases for fixed $a_{0}$ both
approximations tend to grow, so they end up greatly overestimating the exact
result. Hence, for a brief $b_{0}$ range LCF can, in fact, provide a better
approximation than LCF+1. This relates to a well-known behavior of asymptotic
series, for which adding more terms proves helpful in a closer vicinity of the
limit, but counterproductive further away from it.

We must keep in mind that, for each of the terms forming $\mathbb{P}\left(  s\right)  $, the dependence on $b_{0}$ appears through the frequencies found in
the exponential, such as $r_{1}/\left(  2b_{0}\right)  $. Therefore, the
closeness to the LCF limit also depends on $s$, i.e. on the momenta (cf.~\cite{Dinu:2015aci,DiPiazza:2017raw,Podszus:2018hnz,Ilderton:2019kqp}). The
larger the frequencies, the closer we are to that limit. Looking at them we
see that in the vicinity of the point $s_{3}=1$ we are closer to LCF than near
the points $s_{1}=1$ or $s_{2}=1$. It is near the latter points that
$\mathbb{P}_{\rm dir}^{22\rightarrow1}\left(  s\right)  $ starts to first grow as $b_{0}$
is increased, eventually even dominating $\mathbb{P}_{\rm dir}^{22\rightarrow2}\left(
s\right)  $.\ As $b_{0}$ becomes large, almost all of $\mathbb{P}\left(
s\right)$ comes
from the narrow peaks $\mathbb{P}_{\rm dir}^{22\rightarrow1}\left(  s\right)  $ has near
$s_{1}=1$ or $s_{2}=1$, a fact also due to the presence of the factor
$q_{1}^{-2}$ in the integral. 
To study the dominant contribution at large $b_0$ and $a_0\sim1$ we can neglect both the two-step term $\mathbb{P}_{\rm dir}^{22\to2}$ and the first term in the division $I=$ $I^{aa}$ $+$ $I^{a0}+I^{0a}$ mentioned in Appendix~\ref{Numerical methods}, so we only have to compute two nested integrals, which makes this an easy task. All one-step terms grow relative to the two-step, including the much harder to compute  $\mathbb{P}_{\rm ex}^{22}$, but  $\mathbb{P}_{\rm dir}^{22\to1}$ dominates globally, so at large $b_0$ we may neglect $\mathbb{P}_{\rm ex}^{22}$ when computing the probability. The next-to-leading term will be $\mathbb{P}_{\rm dir}^{11}$, which is even faster to compute. 
A negligible two-step term will also occur if $b_0$ is not large, but $a_0$ is small. However, in that case we have no reason to neglect $\mathbb{P}_{\rm ex}^{22}$, which can be comparable to the other terms, as seen in Fig.~\ref{perturbativeplot}.

The $\mathbb{P}_{\rm dir}^{22\to1}$ dominance is also interesting from a conceptual point of view, because, together with the two--step $\mathbb{P}_{\rm dir}^{22\to2}$ it can be obtained from our gluing approach~\cite{Dinu:2018efz,gluingPaper}. The only difference consists in what step-function combination is put in the lightfront time integrand, see~\eqref{StepsForStepsSep}. So the gluing technique provides two things:
On the one hand it provides a beyond-LCF generalization of what we call the N-step part (the cascade part) of tree-level diagrams with N final state particles.
This is done through a top-down treatment of the spin/polarization of the intermediate particles. On the other hand, we also provide a large part of the rest of the probability. How this compares with the remaining terms depends on the process and on the different parameters involved. Knowing that for the trident process in a long pulse, $a_{0}\ge 1$ and any $b_{0}$ the sum of the two terms, $\mathbb{P}_{\rm dir}^{22}$, is dominating the others is also important for future generalizations to higher order processes.

When at least one of the frequencies becomes too
large, the result is exponentially suppressed. Due to this the $\mathbb{P}\left(
s\right)  $ distribution covers a small region at the centre of the $s$
triangle for low $b_{0}$, which expands more and more to the sides as $b_{0}$
grows. For $a_{0}>1$ the extent of this region is essentially controlled by
the value of the $\chi$ parameter, as for the LCF approximation. But notice
that a small characteristic frequency also leads to a reduction in probability
density. This explains the depression found in the middle of the triangular
distributions found at the lower left corner of Fig.~\ref{3Dlin}. It may also be a good idea to look slantwise at Fig.~\ref{bothSectionsFig}~ and~\ref{LongPulsePlot}, to compare the plots that correspond to the same $b_{0}$ values. In this way we do not risk to overestimate the speed of convergence to LCF as the intensity (proportional to  $a_{0}^{2}$) is increased at fixed collision energy. 

\begin{figure}
\includegraphics[width=\linewidth]{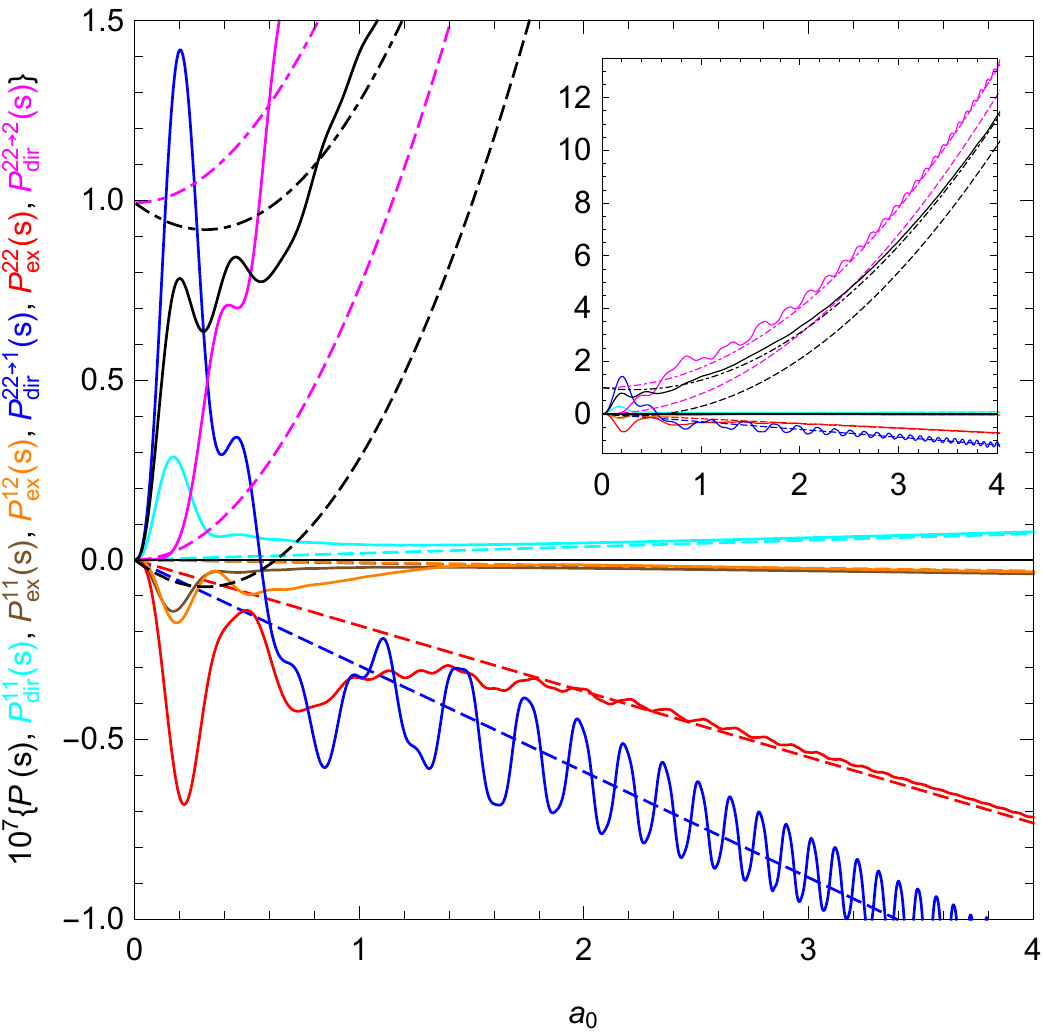}
\caption{Dependence on $a_0$ for $s_1=s_2=s_3=1/3$, $\chi=1$ and $\mathcal{T}=\pi$.}
\label{fixedsplot}
\end{figure}

Fig.~\ref{fixedsplot} illustrates, for the short pulse, the transition form the one-step dominated regime to the two-step dominated one, by varying $a_{0}$ at fixed $\chi$ for a symmetric distribution of the lightfront momentum between the three particles. Notice, though, that for subunitary $a_{0}$, the comparison with LCF and, thus, the $\chi$ parameter lose their relevance. It is best
to consider the dependence on $a_{0}$ at constant $b_{0}$.
If $b_{0}$ is large, a small $\chi$ no longer implies an
exponential suppression of the probability. While at large $a_{0}$ and
constant $\chi$, $\mathbb{P}_{2}$ dominates $\mathbb{P}_{1}$ as $O\left(  a_{0}^2\right)
/O\left(  a_{0}\right)$, at low $a_{0}$ and constant $b_{0}$ the
opposite is true, as a direct Taylor expansion of the integrand shows that
$\mathbb{P}_{2}\left(  s\right)  =O\left(  a_{0}^{4}\right)  $ and $\mathbb{P}_{1}\left(
s\right)  =O\left(  a_{0}^{2}\right)  $. Fig.~\ref{perturbativeplot} shows us, for the short pulse, how the two-step term loses importance compared to the one-step term, as $a_{0}$ decreases. All plots for a given $b_{0}$ have the same scale after dividing by $a_{0}^{2}$, so all the one-step curves and the one for the total result tend to a definite limit, while the two-step curve tends to zero, as $O(a_{0}^{2})$. Thus $\mathbb{P}\left(
s\right)  \approx  a_{0}^{2} P^{0}\left(
s\right) $ at low laser intensities. The limit $ P^{0}\left(
s\right) $ is independent of $a_{0}$. For a monochromatic field there is a threshold at $b_0=4$. For the pulse we consider here $P^{0}$ decays exponentially as $b_0<4$ decreases.

The $|\mathbb{P}_{1}\left(  s\right)|/\mathbb{P}_{2}\left(  s\right)$ ratio
decreases not only with $a_{0}$, but also with pulse length. However, if the polarization is
linear, the aforementioned oscillations make this decrease significantly
slower for $\mathbb{P}_{\rm dir}^{22\rightarrow1}\left(  s\right)  $ than for the other
terms making up $\mathbb{P}_{1}\left(  s\right)  $. Only after one averages/integrates
over $s$, they may become comparable. Pulse length also influences the validity of the aforementioned  low $a_{0}$  expansions. The bigger it is, the smaller $a_{0}$ has to be for such perturbative expansions to work well. 
In fact, while an experimentalist's biggest challenge may be to compress a pulse as much as possible, to make it both ultra-short and ultra-intense, the computational challenge is made higher by a longer, more diluted pulse of subunitary $a_{0}$, where all terms, two-step and one-step alike, are comparable and neither a perturbative nor an LCF approach can help.

The $\mathbb{P}_{\rm ex}^{22}\left(  s\right)  $ and $\mathbb{P}_{\rm dir}^{11}\left(  s\right)  $ terms are the closest to LCF, giving a decent order-of-magnitude estimate even at
higher $b_{0}$ values where $\mathbb{P}_{\rm dir}^{22}$ is completely off the mark. A heuristic
explanation is again related to the characteristic frequencies. For large
$b_{0}$, $r_{1}/\left(  2b_{0}\right)  $ can be quite small in some relevant
region, but is added to the larger $r_{2}/\left(  2b_{0}\right)  $ for
$\mathbb{P}_{\rm dir}^{11}\left(  s\right)  \,$, while for $\mathbb{P}_{\rm ex}^{22}\left(  s\right)  $ there are larger frequency oscillations that couple all phase points $\phi_{i}$.

The interference terms $\mathbb{P}_{\rm ex}^{12}\left(  s\right)  $ and $\mathbb{P}_{\rm dir}%
^{12}\left(  s\right)  $ are seen to be relatively small for $a_{0}\geq1$, but
can be the same order of magnitude with the other terms for smaller $a_{0}$.

\section{Numerical methods}\label{Numerical methods intro}

To efficiently compute all terms that make up the exact\ (non-LCF) probability
density $\mathbb{P}(s)$ we have used two different approaches,
benchmarking one against the other when possible. The two classes of numerical
methods have already been mentioned in~\cite{Dinu:2017uoj}, where, however, only class
B was used and only to compute rates, not total probabilities. We now add to
the B-type methods an ingredient to boost computation speed, which proves
vital for the difficult $\mathbb{P}_{\rm ex}^{22}(s)$ term.

We give here a general description of the ideas behind the two classes of
methods. In Appendix~\ref{Numerical methods} we explain in detail the proper implementation of
method $A$ for the most important terms. We stress the fact that, in talking
about regularization, we do not mean the formulas from~\cite{Dinu:2017uoj} are not
well defined as such, but only that, for practical computations, an
infinitesimal epsilon is not convenient.

A. (${\rm A}_1$ and ${\rm A}_2$) This approach is well suited for computing all terms apart from
$\mathbb{P}_{\rm ex}^{22}(s)$ and, in particular, it is vital for an
efficient calculation of $\mathbb{P}_{\rm dir}^{22}(s)$ for realistic laser
pulses, which have many oscillations. While being a bit involved, its
complexity pays off greatly. The main idea is to change variables to the
quantities $\Theta_{ij}$, whose linear combination appears in the exponent of
the integrands, so as to reduce to a minimum the number of fast oscillating
integrals and write the spectrum in the form of a linear combinations of
Fourier transforms plus some residual terms involving lower dimension
quadratures and special functions.

As already mentioned in~\cite{Dinu:2017uoj} the residual terms are computation-friendly
analogues of half-free terms that involve interaction with the background
field at just one of the two steps. They are due to the Heaviside functions
originating in amplitude-level time ordering and do not occur in the two-step
term $\mathbb{P}_{\rm dir}^{22\rightarrow2}(s)$, where the domain of the
$\Theta_{ij}$ integrals has been extended to the whole real axis, or in the
lightfront instantaneous part $\mathbb{P}^{11}(s)$, where there is no
step function.

For $\mathbb{P}_{\rm dir}^{22}(s)$, a manipulation of the Heaviside
functions allows us to obtain triangular domains for the inner double quadrature
in variables ($\phi_{1}$,$\phi_{3}$), which precedes the $(\Theta
_{21},\Theta_{43})$ Fourier transforms, and have integrands made up of
independent factors that can be sampled independently on a grid, thus keeping
the numerical complexity of the ($\phi_{1}$,$\phi_{3}$) quadrature close to
that required by just one single integral. Then, instead of performing the
outer $(\Theta_{21},\Theta_{43})$ integrals as they are, the total
phase of the complex exponential is used in method ${\rm A}_1$ as a new variable for a
last, Fourier, integral, such that the quadrature that precedes it does not
take oscillations from the complex exponential, so it is faster to perform. A specialized routine is used for the last integral, which only approximates the
prefactor and uses exact integration of the resulting polynomial multiplied by
trigonometric functions, so very frequent oscillations in the latter can
accurately be followed without the need for a huge grid.

An even more efficient variant of this method that we will term ${\rm A}_2$, as opposed
to the direct integration ${\rm A}_1$, involves computing intermediate interpolation
tables. These are meant to store, for a given pulse shape and intensity, a
table of partial derivatives of the inner integrals and then their Fourier
transforms, which can be reused to quickly generate results for any
combination of initial/final lightfront momenta ($b_{0}$ and $s$). To use
an ordinary Fourier transform we take advantage of the fact that for the
variable pair $\left(  \Theta_{21},\Theta_{43}\right)  $, too, the integration
domain is triangular and that it can be brought to the upper half-plane by a
45 degree rotation. Moreover, the fact that an equally spaced rectangular grid
in the new variables is a subset of a similar grid for the unrotated
coordinates means we can produce the interpolation tables for the Fourier
integrand efficiently by an independent sampling of the aforementioned factors
and their derivatives in variables $\Theta_{21}$ and $\Theta_{43}$, too. All
these reasons make it a very efficient method, well suited for computing
results for long pulses, where ${\rm A}_1$ will be slow even for the computation of
just a few points. The region covered by the tables has a limited extent, so
an asymptotic expansion of the integrand outside them has to be performed and
its integral can be computed analytically or by complex deformation for fast convergence.

B. This method uses regularization by deformation of the complex integrals
into the complex plane, using the finite epsilon prescription described in~\cite{Dinu:2017uoj}. This is essential for $\mathbb{P}_{\rm ex}^{22}(s)$, for which
the complexity of the formula and its singularities make regularization by
either subtraction of simplified singular terms or partial integration rather
unwieldy for practical purposes. Had the undeformed expression of $\mathbb{P}_{\rm ex}^{22}(s)$~\eqref{P22exGen} been regular or easy to make so, an A-type
method involving changing variables to the total phase and thus limiting fast
oscillations to only one integral would have been convenient.

At first sight, it seems that to produce even just one $\mathbb{P}_{\rm ex}^{22}(s)$ point, for a short pulse length, one needs a huge
computing power. That is because the formula~\eqref{P22exGen} involves four nested integrals, with long oscillating tails
towards infinity, and the integrand itself has a complex formula. \ When
counting integrals we do not include the averages $\langle a\rangle$ and $\langle a^2\rangle$ which we pre-compute and tabulate
for all terms and methods.

A great boost in speed can be obtained, for both exact and LCF results, by
changing variables from $\phi_{i}\,\ $(for probabilities) or $\theta_{ij}$
(coordinates relative to a given point, for rates) to one radial and some
angular coordinates, depending on the dimensionality $d$ of the quadrature.
Assuming the peak or center of the pulse is at the origin, we use polar
coordinates for $\mathbb{P}^{11}(s)$~\eqref{P11Gen}, spherical
coordinates for $\mathbb{P}^{12}(s)$~\eqref{P12dirGen},~\eqref{P12exGen}, given by
\be
\phi_{1}=r\sin\theta\cos\beta \quad \phi_{2}=r\cos\theta \quad \phi_{3}=r\sin\theta
\sin\beta
\ee
with $r\in[0,\infty)$, $\theta\in[  0,\pi]$ and
$\beta\in[\pi/4,5\pi/4]$, to implement time ordering, while for
$\mathbb{P}_{\rm ex}^{22}(s)$~\eqref{P22exGen} we use a custom type of
generalized spherical coordinates:
\be
\begin{split}
	&\phi_{1}=r\cos\theta\cos\beta \quad \phi_{2}=r\sin\theta\cos\gamma \\
	&\phi_{3}=r\cos\theta\sin\beta \quad \phi_{4}=r\sin\theta\sin\gamma \;,
\end{split}
\ee
where $r\in\left[  0,\infty\right)$,  $\theta\in\left[  0,\pi/2\right]  $ and
$\beta,\gamma\in\left[  \pi/4,5\pi/4\right]  $.

For computing the rate at a phase point $\phi$ a similar procedure is very
useful, but using spherical coordinates relative to that point for
$\mathcal{R}_{\rm ex}^{22}\left(  s\right)  $ and polar ones for $\mathcal{R}%
^{12}\left(  s\right)  $.

\bigskip\ The factor $r^{d-1}$ in the Jacobian associated with this change
amplifies the asymptotic oscillations and makes integrating to infinity
unfeasible, in particular for $\mathbb{P}_{\rm ex}^{22}\left(  s\right)  $, where
$d=4$. But, if one implements a cut-off in the radial direction and varies it,
one sees that convergence of the total result within a very good precision is
quickly reached. The reason is that if one first integrates on a hypersphere,
at constant $r$, due to the many canceling oscillations on its surface, one
obtains a result that decreases fast asymptotically, a thing unaffected by the
$r^{d-1}$-proportional dilation of the hypersphere, as the density of the surface oscillations per unit solid angle also increases. After the integration over the radial
direction and $\theta$ angular direction and with an appropriate choice of
$\varepsilon$, the remaining angular integrals will have a limited amount of
oscillations in them and few, if any, sign changes, which makes their
computation efficient and accurate.

We want to limit the oscillations in the angular directions to a minimum and
avoid small but fast canceling oscillations coming from the boundary of the
ball, centered on the pulse, to which we restrict the integral. Therefore,
instead of an abrupt cut-off at the boundary of the ball, we use a mollified
one, multiplying the integrand by a $C^{\infty}$ function of $r$ that equals
unity in this ball and slowly and smoothly goes to zero between it and a
larger, concentric ball. In this way, convergence when varying the cut-off
radius is faster and the computation time for a given radius is reduced.

We obtain a fast convergence even for small $\chi$ values, such as $1/2$,
which are in the region of exponential suppression, unless only $a_0$ is small and not $b_0$. In this limit it is a
greater challenge to compute the integral of what are initially fast
oscillations, whose almost perfect cancelling brings the result down by
several orders of magnitude, so the advantages of our method are even more apparent.

Especially for small $a_{0}$, it is very helpful to subtract from the
integrand its free-field ($a_{0}=0$) counterpart, since the latter can be
proven to yield a vanishing integral, as expected. We checked that in doing so
the change in the result is negligible, providing another good test for our method.

We have also checked that, for $\mathbb{P}_{11}$ and $\mathbb{P}_{12}$, methods A and B give
consistent results. We also checked that the exact and LCF trident rates, like
the ones shown in~\cite{Dinu:2017uoj}, can be obtained more efficiently using spherical coordinates.

Outside the LCF approximation the rates are oscillating many times while
slowly decaying outside the pulse, so it is preferable to apply method B
directly to the probability terms. For LCF calculations the complex
deformation introduces rapid decay at infinity, so the radial integrals are
fast converging without any cut-off and the use of rates, tabulated
in advance as function of $\chi$, is a good idea.

\section{Conclusions}

We have studied the trident process in laser pulses with $a_0\gtrsim1$, which mean that we have gone beyond the LCF approximation. We have found that for $a_0\gtrsim1$ the (longitudinal) momentum spectrum has a richer structure compared to LCF.
The corrections to the two-step, which we call one-step terms, are difficult to compute for long pulses, but they are anyway expected to be small for sufficiently long pulses. For shorter pulses they can be important though. We have studied all contributions for a short pulse, and found that the one-step terms can even become larger than the two-step for large electron momentum $b_0$. That the one-step can become dominant is expected from the perturbative limit, because only the one-step contributes to leading order $\mathcal{O}(a_0^2)$. 
This is also consistent with the fact that the convergence to the LCF approximation needs larger $a_0$ for larger $b_0$.       

For a field with longer pulse length we have explicitly demonstrated that, unless $b_0$ is large or $a_0$ small, the one-step terms become negligible and so the two-step can indeed be used to approximate the probability. This is evidence in support of using our new gluing approximation~\cite{Dinu:2018efz,gluingPaper} for studying other higher-order processes in long laser pulses with moderately high intensity $a_0\gtrsim1$.

\acknowledgments
We thank Christian Kohlf\"urst and Andre Sternbeck for advice on the use of computer clusters.
We thank Sebastian Meuren for discussions about the planned experiments at FACET-II. 
We thank Tom Blackburn, Anton Ilderton and Anthony Hartin for useful discussions about the field shape.  
V.~Dinu is supported by the 111 project 29 ELI RO financed by the Institute of Atomic
Physics A, and G.~Torgrimsson was supported by the Alexander von Humboldt foundation during most of the work and writing of this paper.

\appendix

\appendix

\section{Numerical methods}\label{Numerical methods}

Let us study the application of methods ${\rm A}_1$ and ${\rm A}_2$ to the integral $I$ appearing
in the expression:
\begin{equation}\label{pedir}
\mathbb{P}_{\rm dir}^{22}\left(  s\right)  =-\frac{\alpha^{2}}{4\pi
	^{2}b_{0}^{2}}\left[  \frac{I}{q_{1}^{2}}+\left(  1\leftrightarrow2\right)
\right] \;.
\end{equation}
In order to make the changes of variables $\theta_{21}\rightarrow\Theta_{21}$,
$\theta_{43}\rightarrow\Theta_{43}$ we notice the very useful fact that, since
$\Theta_{ij}$ always grows with $\phi_{i}\,$\ and decreases with $\phi_{j}$,
we have the implications:
\be
\begin{split}
	\left(  \theta_{42}>0,~\Theta_{43}<\Theta_{21}\right)   &  \Rightarrow
	\theta_{31}>0\\
	\left(  \theta_{31}>0,~\Theta_{43}>\Theta_{21}\right)   &  \Rightarrow
	\theta_{42}>0 \;,
\end{split}
\ee
so we can express the time-ordering step functions as:
\be\begin{split}
	\theta\left(\theta_{42}\right)  \theta\left(\theta_{31}\right)
	=&\theta\left(\theta_{42}\right)  \theta\left(\Theta_{21}-\Theta
	_{43}\right) \\
	&+\theta\left(  \theta_{31}\right)\theta\left(\Theta
	_{43}-\Theta_{21}\right) \;.
\end{split}
\ee
Therefore, we break the integral into two pieces, for which we choose the new
variables to be $\left(  \phi_{1},\Theta_{21},\phi_{3},\Theta_{43}\right)  $
and $\left(  \phi_{2},\Theta_{21},\phi_{4},\Theta_{43}\right)  $,
respectively. At first sight, the pieces are complex conjugates, so it seems we only have to compute
the first one. The integration domain determined by the step functions is now
triangular for both the pairs $\left(  \phi_{1},\phi_{3}\right)  $ and
$\left(  \Theta_{21},\Theta_{43}\right)  $. Hence, we would like to be able to put the integral into the form:%
\begin{equation}
I=\int \ud\Theta_{21}\int_{\Theta_{21}}\ud\Theta_{43}\int \ud\phi_{1}\int_{\phi_{1}%
}\ud\phi_{3}\operatorname{Re}J \label{oneint}%
\end{equation}

We will however see later on that for some terms the two apparently complex conjugate integrals need to be kept separate and the way their boundary values extend to infinity must be correlated in a particular way, corresponding to a symmetric integration in the initial variables. 
To regularize the expression, i.e. be able to obtain a finite integrand $J$
without further need of the epsilon prescription, we keep in mind that $J$ is
a linear combination of factorized terms of the form $F_{i}\left(  \phi
_{1},\Theta_{21};\omega_{1}\right)  F_{j}\left(  \phi_{3},\Theta_{43}%
;\omega_{2}\right)  $, where $\omega_{1}=\frac{r_{1}}{2b_{0}}$, $\omega_{2}=\frac{r_{2}}{2b_{0}}$,
\be
F_{i}\left(  \phi,\Theta;\omega\right)  =f_{i}\left(  \phi,\Theta\right)
e^{i\omega\Theta} \;,
\ee
\be
f_{0}=\tfrac{1}{\theta^{2}\left(  \partial\Theta/\partial\theta\right)
	_{\phi}},~f_{1}=\tfrac{1}{\theta\left(  \partial\Theta/\partial\theta\right)
	_{\phi}} \;,
\ee
and
\be
\begin{split}
	&\left\{  f_{2},f_{3},f_{4},f_{5}\right\} \\
	&=\tfrac{\left\{  \mathbf{w}%
		_{1}\cdot\mathbf{w}_{2},w_{11}w_{22}-w_{12}w_{21},w_{11}w_{21}-w_{12}%
		w_{22},w_{12}w_{21}+w_{11}w_{22}\right\}  }{\theta\left(  \partial
		\Theta/\partial\theta\right)  _{\phi}} \;.
\end{split}
\ee
Only $F_{0}$ and $F_{1}\,$are singular functions at $\Theta=0$, from whom we
plan to subtract simpler, analytically integrable functions $F_{i}^{0}$
possessing the same singular part, using the decomposition
\be\label{deco}
F_{i}F_{j}=G_{i}G_{j}+G_{i}F_{j}^{0}+F_{i}^{0}G_{j}+F_{i}^{0}F_{j}^{0} \;,
\ee
\be
G_{i}=F_{i}-F_{i}^{0},
\ee
and making sure that the last term in \ref{deco} vanishes when integrated.

As explained in~\cite{Dinu:2017uoj}, doing this
before the change of variables to $\Theta_{ij}$ and using for $F_{i}^{0}$ the
free field $\left(  a_{0}=0\right)  $ version of $F_{i}$, provides an
interpretation of the second and third terms in \ref{deco} as half-virtual, that is involving an
interaction with the field at only one of the two steps. From a computational point of
view, though, it is preferable to have only one Fourier exponential as a
common factor, so as to regularize directly the pre-exponential factor. Hence
we choose:
\be
F_{i}^{0}=f_{i}^{0}e^{i\omega\Theta};~f_{0}^{0}=\tfrac{1}{\Theta^{2}}%
;~f_{1}^{0}=\tfrac{1}{\Theta};~f_{i}^{0}=0,~i>1,
\ee
and write all formulas in terms of the regularized factors
\be
g_{i}=f_{i}-f_{i}^{0}.
\ee

Successively
plugging into the linear combination $J$ just one of the first three terms in
\ref{deco} produces the decomposition:
\be
I=I^{aa}+I^{a0}+I^{0a} \;.
\ee
We detail them below, starting with the first, four dimensional quadrature term:%
\begin{equation}\label{i12}
I^{aa}=\operatorname{Re}\int \ud\Theta_{21}\int_{\Theta_{21}}\ud\Theta_{43}\left(
A+iB\right)e^{i\left(  \omega_{1}\Theta_{21}+\omega_{2}%
	\Theta_{43}\right)}  \;,
\end{equation}
where
\be\begin{split}
	A&=T_{00}+\tfrac{\kappa_{01}}{2}T_{10}+\tfrac{\kappa_{23}}{2}%
	T_{01}+\tfrac{\kappa_{01}}{2}\tfrac{\kappa_{23}}{2}\left(  T_{11}%
	+\tfrac{T_{22}}{\omega_{1}\omega_{2}}\right) \\
	B&  =\tfrac{\kappa_{01}}{2\omega_{1}}\left(  T_{20}+\tfrac{\kappa_{23}%
	}{2}T_{21}\right)  +\tfrac{\kappa_{23}}{2\omega_{2}}\left(  T_{02}%
	+\tfrac{\kappa_{01}}{2}T_{12}\right)  
\end{split}
\ee
and $T_{ij}$ are obtained from double integrations which can be done
efficiently, as they are defined on a triangular domain and the oscillations
of the integrand only come from the pulse shape:%
\be
\begin{split}
	T_{00}&=-S_{11}-S_{44}-S_{55} \quad T_{10}=S_{61},~T_{01}=-S_{16} \\
	T_{11}&=S_{66}+S_{33} \quad T_{22}=-S_{00} \quad T_{20}=-S_{01} \\
	T_{21}&=-S_{06} \quad T_{02}=S_{10} \quad T_{12}=-S_{60} \;,
\end{split}
\ee
where
\be
S_{ij}=\int\ud\phi_{1}\int_{\phi_{1}}\ud\phi_{3}g_{i}\left(\phi_{1}%
,\Theta_{21}\right)  g_{j}\left(  \phi_{3},\Theta_{43}\right)
\ee
and $g_{6}=g_{1}+g_{2}$.

A special routine was written for computing $S_{ij}$. For method ${\rm A}_1$ it first uses a preliminary
adaptive Simpson method to sample the two functions\thinspace\ $g_{i}$ and
$g_{j}$ on suitable nonuniform grids. Then follows a $\phi_{1}-$integration of the product of the piecewise
polynomial approximations generated for $g_{i}\left(  \phi_{1},\Theta
_{21}\right)  $ and $\int_{\phi_{1}}\ud\phi_{3}g_{j}\left(  \phi_{3},\Theta
_{43}\right)  $. In this way we only sample the two functions separately on a
line, as for a one-dimensional integral. For method  ${\rm A}_2$  each of the functions $g_{i}$  is accompanied in the above procedure by a few of its $\Theta_{ij}$ derivatives and, thus, a whole matrix of partial derivatives of $S_{ij}$ with respect to $(\Theta_{21},\Theta_{43})$ is produced. This is because method ${\rm A}_1$ is aimed at producing just one point value, while method ${\rm A}_2$ is meant to produce a whole set of
spectra, for any $b_{0}$ and $(s_{1},s_{2})$ values.

For method ${\rm A}_1$ it is best to change variable to the exponent $\Theta=\omega
_{1}\Theta_{21}+\omega_{2}\Theta_{43}$ and leave it for the last Fourier integral.

The procedure for ${\rm A}_2$ is more complex, as we want to create tables of values
for \ref{i12} and we start by tabulating $T_{ij}$ together with a matrix of partial derivatives, obtained as described above. To turn the integration domain into the upper half-plane, we make
a 45 degree rotation by changing the variables to $Z_{1}=\frac{\Theta
	_{43}+\Theta_{21}}{2}$ and $Z_{2}=\frac{\Theta_{43}-\Theta_{21}}{2}$. We then compute Fourier transforms with new frequencies
$o_{1/2}=\omega_{1}\pm\omega_{2}$ for each element of the matrix of partial derivatives of each term $T_{ij}$ and a dense set of
$o_{1}$ and $o_{2}$ values so as to create up to 9 interpolation tables, for
general polarization.

The factor $1/q_{1}^{2}$ in \ref{pedir} can ruin precision in the vicinity of
$s_{1}=1$, but this can be avoided by a double partial integration, writing:
\be
I^{aa}=-\frac{2}{o_{1}^{2}}\operatorname{Re}\int \ud Z_{1}\int_{0}
\ud Z_{2}\left(  \bar{A}+i\bar{B}\right)  e^{i\left(  o_{1}Z_{1}
	+o_{2}Z_{2}\right)} \;,
\ee
where $\bar{A}=\frac{\partial^{2}A}{\partial Z_{1}^{2}},~\bar
{B}=\frac{\partial^{2}B}{\partial Z_{1}^{2}}$ are linear combinations of $\bar{T}_{ij}=\frac{\partial^{2}T_{ij}}{\partial Z_{1}^{2}}$. The replacement of $T_{ij}$ by $\bar{T}_{ij}$ means that, for a given
maximum total order of the $Z$ partial derivatives, we now need to add two more orders in the
tables that preceed the 45 degree rotations, whose elements are
linear combinations of:
\be
S_{ij}^{(pq)}=\int \ud\phi_{1}\int_{\phi_{1}}\ud\phi_{3}g_{i}^{\left(  p\right)
}\left(  \phi_{1},\Theta_{21}\right)  g_{j}^{\left(  q\right)  }\left(
\phi_{3},\Theta_{43}\right)  ,
\ee
where $g_{i}^{\left(  p\right)  }\left(  \phi,\Theta\right)  =\partial
^{p}g_{i}\left(  \phi,\Theta\right)  /\partial\Theta^{p}$. We tabulate the set
of partial derivatives on a grid with step $h$, so that $Z_{1}=j_{1}h$,
$Z_{2}=j_{2}h$, $j_{2}\geq0$ hence $\Theta_{43}=i_{1}h,~\Theta_{21}=i_{2}h$
and $i_{1/2}=j_{1}\pm j_{2}$. This can be done highly efficiently in the
following manner:

In order to reduce memory storage, we concentrate at one time on just one small piece of the $\phi_{1}$ integrals,
covering an interval of length $l$, adding to the contribution of the previous intervals. Say we use derivatives up to order 4.
For a grid of $i_{1}$ and $i_{2}$ values we determine, using an adaptive
Simpson method, piecewise polynomial approximations for $g_{i}^{\left(p\right)}\left(\phi_{1},\Theta_{21}\right)  $ and $\int_{\phi_{1}}
\ud\phi_{3}g_{j}^{\left(  q\right)  }\left(  \phi_{3},\Theta_{43}\right)  $ on
two different grids of $\phi_{1}$ values, for $p,q=0,1,...,6$ and for a
square $n\times n$ matrix of $\left(  \Theta_{21},\Theta_{43}\right)  $
values, which cover only a small square part of the domain we want covered.

If $n$ is large enough, the computation of the functions and derivatives, the
adaptive Simpson procedure and the subsequent sorting of the grid points cost
little compared to the $n^{2}$ $\phi_{1}$ integrals that are performed, using
a custom integration tool for the product of piecewise polynomial
interpolations. Then the procedure is repeated for all relevant $\phi_{1}$
intervals and, after implementing the rotation, we save to an external memory
the values of $\bar{T}_{ij}$ and its derivatives on a square matrix of equally
spaced $\left(  Z_{1},Z_{2}\right)  $ values, one out of many that are to be
computed and stored for the next step. If the grid spacing$\ h$ is small
enough, by a Taylor expansion, we can find with good precision $\bar{T}_{ij}$
at any point inside the square block where the $\left(Z_{1},Z_{2}\right)  $
values lie.

The second step involves a custom code for implementing the double Fourier
transform, using, for each $h$ interval centered at a grid point, analytical
formulas for the integral of an oscillating exponential multiplied by the
aforementioned Taylor series. Since we cannot compute an infinite number of
blocks, we cover a finite rectangle with these blocks and use for $\bar{T}%
_{ij}$ an asymptotic extrapolation as a polynomial decay times oscillation,
which can be used to extend the integration to infinity, with good
approximation. In this asymptotic region the integral can be expressed in
terms of trigonometric integrals, or a complex rotation of
the integration path may be used to turn the oscillating exponential into a decaying one.

The double Fourier transform of each $\bar{T}_{ij}$ is computed and stored as
a table for a dense set of $o_{1}$, $o_{2}$ values, with $\left\vert
o_{2}\right\vert <o_{1}<o_{\max}$, where the $o_{\max}$ is chosen large enough
to provide all the nonnegligible part of the spectrum for the minimum $\chi$ value we have in mind. Then,
for any $b_{0}$ and $s$ values, $I^{aa}$ is produced from interpolations of
the data in the tables. Next we describe the residual terms, choosing $I^{a0}$
for more detailed explanations.

Let $\tilde{g}_{i}$ be the function obtained from $g_{i}$ by replacing
$\phi_{1}$ by $\phi_{2}$ as variable independent of $\Theta_{21}$.
\begin{widetext}
	\be
	\begin{split}
		I^{a0}=&\int \ud\Theta_{21}\int_{\Theta_{21}}\ud\Theta_{43}\int \ud\phi_{1}\int
		_{\phi_{1}}^{L_{1}\rightarrow\infty}\ud\phi_{3}\left(  \frac{-id_{0}}%
		{\Theta_{43}^{2}}+\frac{d_{1}}{\Theta_{43}}\right)  e^{i\omega_{2}\Theta_{43}%
		} \\
		&\times\left(  ic_{0}g_{0}\left(  \phi_{1},\Theta_{21}\right)  +c_{1}%
		g_{1}\left(  \phi_{1},\Theta_{21}\right)  +c_{2}g_{2}\left(  \phi_{1}%
		,\Theta_{21}\right)  \right)  e^{i\omega_{1}\Theta_{21}} \\
		& +\int \ud\Theta_{21}\int^{\Theta_{21}}\ud\Theta_{43}\int \ud\phi_{2}\int_{\phi_{2}}^{L_{2}\rightarrow\infty}\ud\phi_{4}\left(  \frac{-id_{0}}{\Theta_{43}^{2}%
		}+\frac{d_{1}}{\Theta_{43}}\right)  e^{i\omega_{2}\Theta_{43}} \\
		&\times\left(  ic_{0}\tilde{g}_{0}\left(  \phi_{2},\Theta_{21}\right)
		+c_{1}\tilde{g}_{1}\left(  \phi_{2},\Theta_{21}\right)  +c_{2}\tilde{g}_{2}\left(
		\phi_{2},\Theta_{21}\right)  \right)  e^{i\omega_{1}\Theta_{21}} \;,
	\end{split}
	\ee
\end{widetext}
where $c_{2}=\frac{\kappa_{01}}{2}$, $c_{1}=c_{2}-1$, $c_{0}=-\frac{2b_{0}}{r_{1}%
}c_{2}$, $d_{2}=\frac{\kappa_{23}}{2}$, $d_{0}=-\frac{2b_{0}}{r_{2}}d_{2}$, and
$d_{1}=d_{2}+1$.

We have to write the expression like this, before going to just one integral
like in \ref{oneint}, for an essential ingredient is finding out how $L_{1}$
and $L_{2}$ are to be related for a symmetric integration around the average phase
$\sigma_{ij}$, instead of a discontinuous choice of the variables that are independent from 
$\Theta_{ij}$. By breaking the integration domain into
two parts, each with a different definition of the variables independent of
$\left(  \Theta_{21},\Theta_{43}\right)  $, we effectively severed the
$\Theta_{21}$ and $\Theta_{43}$ integrals and reattached them with a
displacement, which has to be taken into account. Hence we cannot just keep
$L_{1}$ and $L_{2}$ equal as they tend to infinity, for this would give a value for the
probability that can even be negative at low $a_{0}$. Instead, since for a
fixed $\Theta_{43}$ value, at large $\phi$, $\Theta_{43}=\theta_{43}$, to keep the
integration symmetric, we should choose $L_{1}%
=L-\Theta_{43}/2,~L_{2}=L+\Theta_{43}/2$, with $L\rightarrow\infty$. We have checked that this A-type method gives the same results as what we obtain by integrating the rates in~\cite{Dinu:2017uoj}, which were obtained with a B-type method.

We introduce the following special functions:
\be
\begin{split}
	W\left(  x\right)   &  =\int_{-\infty}^{x}\!\ud y\frac{\exp\left(  iy\right)
	}{y+i\varepsilon}\rightarrow\operatorname{Ei}\left(  ix\right)
	-i\pi\operatorname*{sgn}\left(  x\right) ,
\end{split}
\ee

\be
\begin{split}
	W_{2}\left(  x\right)   =\int_{-\infty}^{x}\!\ud y\frac{\exp\left(  iy\right)
	}{\left(  y+i\varepsilon\right)  ^{2}} \rightarrow iW\left(x\right)  -\frac{e^{ix}}{x}.%
\end{split}
\ee

We use the above integrals, then in the second term we swap the two indices of
$\Theta_{21}$, after which we change the sign of $\Theta_{21}$ to keep the same exponent.
\begin{widetext}
	\be\begin{split}	
		2I^{a0}=&\int \ud\Theta_{21}\int \ud\phi_{1}\left[  \left(  \phi_{1}-L\right)
		\left(  d_{1}W\left(  \omega_{2}\Theta_{21}\right)  -id_{0}\omega_{2}W_{2}\left(
		\omega_{2}\Theta_{21}\right)  \right)  +\frac{1}{2}\left(  \frac
		{d_{1}e^{i\omega_{2}\Theta_{21}}}{i\omega_{2}}-id_{0}W\left(  \omega_{2}%
		\Theta_{21}\right)  \right)  \right] \\
		&\times\left(  -ic_{0}g_{0}\left(  \phi_{1},\Theta_{21}\right)  +c_{1}%
		g_{1}\left(  \phi_{1},\Theta_{21}\right)  +c_{2}g_{2}\left(  \phi_{1}%
		,\Theta_{21}\right)  \right)  e^{i\omega_{1}\Theta_{21}} \\
		&+\int \ud\Theta_{21}\int \ud\phi_{1}\left[  \left(  L-\phi_{1}\right)  \left(
		d_{1}W\left(  \omega_{2}\Theta_{21}\right)  -id_{0}\omega_{2}W_{2}\left(
		\omega_{2}\Theta_{21}\right)  \right)  +\frac{1}{2}\left(  \frac
		{d_{1}e^{i\omega_{2}\Theta_{21}}}{i\omega_{2}}-id_{0}W\left(  \omega_{2}%
		\Theta_{21}\right)  \right)  \right]  \\
		&\times\left(  -ic_{0}\tilde{g}_{0}\left(  \phi_{1},\Theta_{21}\right)
		+c_{1}\tilde{g}_{1}\left(  \phi_{1},\Theta_{21}\right)  +c_{2}\tilde{g}%
		_{2}\left(  \phi_{1},\Theta_{21}\right)  \right)  e^{i\omega_{1}\Theta_{21}} 
	\end{split}
	\ee
	We now use the symmetric role of the two $\phi$ values, which implies that
	$\tilde{g}_{0}\left(  \phi_{1},\Theta_{21}\right)  =g_{0}\left(  \phi
	_{1},-\Theta_{21}\right)  $ and $\tilde{g}_{i}\left(  \phi_{1},\Theta
	_{21}\right)  =-g_{i}\left(  \phi_{1},-\Theta_{21}\right)  $, $i\neq0$.
	
	Using the notation $W_{0}\left(  x\right)  =W\left(  x\right)  e^{-ix}$ for
	what is a non-oscillating function, easy to be produced from an interpolation
	table and series expansions near the origin and at large values, we obtain the final formulas
	for the residual terms:%
	\be
	\begin{split}
		I^{a0} &  =\operatorname{Re}\int_{0}\ud\Theta\exp\left(  i\omega\Theta\right)
		\left[  W_{0}\left(  \omega_{2}\Theta\right)  +\tfrac{id_{0}}{\Theta}\right]
		\left[  c_{1}G_{1}+c_{2}G_{2}-ic_{0}G_{0}\right]  \\
		&  +\operatorname{Im}\int_{0}\ud\Theta\exp\left(  i\omega\Theta\right)  \left[
		\tfrac{d_{1}}{\omega_{2}}+d_{0}W_{0}\left(  \omega_{2}\Theta\right)  \right]
		\left[  c_{1}\check{G}_{1}+c_{2}\check{G}_{2}-ic_{0}\check{G}_{0}\right]  \;,
	\end{split}
	\ee
	\be
	\begin{split}
		I^{0a} &  =\operatorname{Re}\int_{0}\ud\Theta\exp\left(  i\omega\Theta\right)
		\left[  -W_{0}\left(  \omega_{1}\Theta\right)  +\tfrac{ic_{0}}{\Theta}\right]
		\left[  d_{1}G_{1}+d_{2}G_{2}-id_{0}G_{0}\right]  \\
		&  +\operatorname{Im}\int_{0}\ud\Theta\exp\left(  i\omega\Theta\right)  \left[
		\tfrac{c_{1}}{\omega_{1}}+c_{0}W_{0}\left(  \omega_{1}\Theta\right)  \right]
		\left[  d_{1}\check{G}_{1}+d_{2}\check{G}_{2}-id_{0}\check{G}_{0}\right]  \;,
	\end{split}
	\ee
\end{widetext}
where $G_{i}=\int\phi\bar{g}_{i}\left(  \phi,\Theta\right)
\ud\phi,~\check{G}_{i}=\int\check{g}_{i}\left(  \phi,\Theta\right)  \ud\phi$ and
$\bar{g}_{0}\left(  \phi,\Theta\right)  \equiv g_{0}\left(  \phi
,\Theta\right)  -g_{0}\left(  \phi,-\Theta\right)$, $\bar{g}_{i}\left(
\phi,\Theta\right)  \equiv g_{i}\left(  \phi,\Theta\right)  +g_{i}\left(
\phi,-\Theta\right)$ for $i\neq0$,
$\check{g}_{0}\left(  \phi,\Theta\right)  \equiv\frac{g_{0}\left(  \phi
	,\Theta\right)  +g_{0}\left(  \phi,-\Theta\right)  }{2}$, $\check{g}_{i}\left(
\phi,\Theta\right)  \equiv\frac{g_{i}\left(  \phi,\Theta\right)  -g_{i}\left(
	\phi,-\Theta\right)  }{2}$ for $i\neq0$.
We have checked analytically that for $a_0\ll1$ we find agreement between $I^{0a}$ and the high-energy limit of perturbative $\mathcal{O}(a_0^2)$ trident. Since both the Re... and Im... terms in $I^{0a}$ contribute to leading order, this is a nontrivial check of the choice of the boundaries $L_1$ and $L_2$ above.  

Similarly to what has already been described before, method ${\rm A}_1$ uses direct integration, while method ${\rm A}_2$ involves the tabulation of the functions
$G_{i}$,$~\check{G}_{i}$ and a few of their derivatives, for a
large region of $\Theta$ values, so as to be used repeatedly to compute the integrals for any $b_0$ and $s_i$ values. Here, too, a double integration by parts is very useful to prevent precision loss.
We can extend the tables enough as to not need asymptotic expansions, as convergence is fast enough.

Next we want to calculate~\eqref{P12dirGen} and~\eqref{P12exGen}:
\be
\mathbb{P}_{\rm dir/ex}^{12}\left(  s\right)  =\frac{\alpha^{2}}{4\pi^{2}b_{0}%
}\left[  I_{\rm dir/ex}^{12}+\left(  1\leftrightarrow2\right)  \right]
\ee
To make the formulas very similar to the ones before, we interchange
even and odd indices, so $\phi_{1}\leftrightarrow\phi_{2}$, $\phi
_{3}\rightarrow\phi_{4}$ and the three phase points are $\phi_{1},\phi
_{2},\phi_{4}$. The direct term needs no regularization and, therefore, it has
no residual terms,  $I_{\rm dir}^{12}=I_{\rm dir}^{12aa}$, while the exchange one has the decomposition $I_{\rm ex}=I_{\rm ex}^{12aa}+I_{\rm ex}^{12a0}+I_{\rm ex}^{120a}$
\be
\begin{split}
	I_{\rm dir/ex}^{12aa} =\int \ud\Theta_{21}\int_{\Theta_{21}}\ud\Theta_{41}B_{\rm dir/ex}^{12}\sin\left(
	\omega_{1}\Theta_{21}+\omega_{2}\Theta_{41}\right)  \;,
\end{split}
\ee
\be
B_{\rm dir}^{12} =\frac{\left(  1+s_{1}\right)  \left(  s_{2}-s_{3}\right)  }%
{q_{1}^{3}}\int \ud\phi_{1}\mathbf{g}_{a}\left(  \phi_{1},\Theta
_{21}\right)  \cdot\mathbf{g}_{a}\left(  \phi_{1},\Theta_{41}\right) \;,
\ee
\be
\begin{split}
	B_{\rm ex}^{12} &  =\int \ud\phi_{1}\bigg[  -\tfrac{q_{1}}{q_{2}^{2}}g_{1}\left(
	\phi_{1},\Theta_{21}\right)  g_{1}\left(  \phi_{1},\Theta_{41}\right)
	\\
	&\hspace{1cm}+\tfrac{s_{1}s_{3}-s_{2}}{q_{1}q_{2}^{2}}\mathbf{g}_{a}\left(  \phi_{1}%
	,\Theta_{21}\right)  \cdot\mathbf{g}_{a}\left(  \phi_{1},\Theta_{41}\right)
	\bigg]  \;, 
\end{split}
\ee

where
$\mathbf{g}_{a}\left(  \phi,\Theta\right)  =\tfrac{\mathbf{w}_{2}/\theta
}{\left(  \partial\Theta/\partial\theta\right)  _{\phi}}$, and

\be
\begin{split}
	I_{\rm ex}^{12a0}+I_{\rm ex}^{120a}  =-2\tfrac{q_{1}}{q_{2}^{2}}\operatorname{Im}\int
	_{0}&\ud\Theta\exp\left(  i\omega\Theta \right)\check{G}_{1}\left(  \Theta\right) \\
	\times&\left[  W_{0}\left(
	\omega_{1}\Theta\right)  -W_{0}\left(  \omega_{2}\Theta \right)  \right] \;.
\end{split}
\ee

Notice that all residual terms vanish in the LCF
approximation, but they can be essential at low $a_{0}$/large $b_{0}$, where
they dominate.

The procedure to compute just the two-step term, $\mathbb{P}_{\rm dir}^{22\rightarrow2}(s)$, is similar to the one for $\mathbb{P}_{\rm dir}^{22}(s)$, except that we use the
average coordinates $\sigma_{21}=\frac{\phi_{1}+\phi_{2}}{2}$ and $\sigma
_{43}=\frac{\phi_{3}+\phi_{4}}{2}$ as independent variables and there are no residual terms. We will not
present the details, as they are easy to infer, but mention that we need only
up to five tables here, for general polarization, if we make the partial
integration used in Eqs. (34)-(37) in~\cite{Dinu:2017uoj}. Again, a double partial
integration can bring out a factor of $\omega_{2}^{-2}$, to mitigate the
singularity $q_{1}^{-2}$ present in the formula when $s_{1}$ is close to unity.  Asymptotic expansions outside the region covered by the tables are essential for the two-step terms, which have slower asymptotic decay than for the integrals encountered before and even more important for the LCF approximation to the two-step probability that can also be computed by the same method.
To compute  $\mathbb{P}^{11}\left(  s\right)  $ we need similar interpolation tables for the Fourier integrand and its transform, but these are one-dimensional, so require fewer resources. Here, too, the integrand is approximated by an expansion outside the tables.

\begin{widetext}

\section{Formulas}\label{Formulas}

For the readers' convenience we collect the formulas from~\cite{Dinu:2017uoj} in this appendix. The total probability is given by the sum of the following 6 terms:  
\be\label{P11Gen}
\{\mathbb{P}_{\rm dir}^{11}(s),\mathbb{P}_{\rm ex}^{11}(s)\}=\frac{\alpha^2}{\pi^2}\int\!\ud\phi_{12}\left\{\frac{1}{q_1^4}+\frac{1}{q_2^4},-\frac{1}{q_1^2q_2^2}\right\}\frac{-s_0s_1s_2s_3}{(\theta_{21}+i\epsilon)^2}\exp\left\{\frac{i}{2b_0}(r_1+r_2)\Theta_{21}\right\} \;,
\ee
\be\label{P12dirGen}
\mathbb{P}_{\rm dir}^{12}(s)=\text{Re }i\frac{\alpha^2}{4\pi^2b_0}\int\!\ud\phi_{123}\theta(\theta_{31})
\frac{(s_0+s_1)(s_2-s_3)D_{12}}{q_1^3(\theta_{21}+i\epsilon)(\theta_{23}+i\epsilon)}\exp\left\{\frac{i}{2b_0}\left[r_1\Theta_{21}+r_2\Theta_{23}\right]\right\}+(s_1\leftrightarrow s_2) \;,
\ee
\be\label{P12exGen}
\mathbb{P}_{\rm ex}^{12}(s)=\text{Re}\frac{-i\alpha^2}{4\pi^2b_0}\int\!\ud\phi_{123}\theta(\theta_{31})
\frac{q_1^2+[s_0s_2-s_1s_3]D_{12}}{q_1q_2^2(\theta_{21}+i\epsilon)(\theta_{23}+i\epsilon)}\exp\left\{\frac{i}{2b_0}\left[r_1\Theta_{21}+r_2\Theta_{23}\right]\right\}+(s_1\leftrightarrow s_2) \;,
\ee
\be\label{P22dirGen}
\begin{split}
	\mathbb{P}_{\rm dir}^{22}(s)=-\frac{\alpha^2}{8\pi^2b_0^2}&\int\!\ud\phi_{1234}\frac{\theta(\theta_{31})\theta(\theta_{42})}{q_1^2\theta_{21}\theta_{43}}\exp\left\{\frac{i}{2b_0}\left(r_1\Theta_{21}+r_2\Theta_{43}\right)\right\}\left\{\frac{\kappa_{01}\kappa_{23}}{4}W_{1234}+W_{1324}+W_{1423}\right.
	\\
	&\left.+\left[\frac{\kappa_{01}}{2}\left(\frac{2ib_0}{r_1\theta_{21}}+1+D_1\right)-1\right]\left[\frac{\kappa_{23}}{2}\left(\frac{2ib_0}{r_2\theta_{43}}+1+D_2\right)+1\right]-D_1D_2\right\}+(s_1\leftrightarrow s_2) \;,
\end{split}
\ee
where $\theta_{ij}=\phi_i-\phi_j$, $\ud\phi_{1234}=\ud\phi_1...\ud\phi_4$,  $r_1=(1/s_1)-(1/s_0)$, $r_2=(1/s_2)+(1/s_3)$,
$\kappa_{ij}=(s_i/s_j)+(s_j/s_i)$, $\Theta_{ij}:=\theta_{ij}M^2_{ij}$, the effective mass~\cite{Kibble:1975vz}
$M^2=1+\langle a_\LCperp^2\rangle-\langle a_\LCperp\rangle^2$, $\langle a\rangle=\frac{1}{\theta}\int_{\phi_1}^{\phi_2}a$,
$D_{12}={\bf \Delta}_{12}\!\cdot\!{\bf \Delta}_{32}$ and
\be\label{Wijkldef}
W_{ijkl}:=(\mathbf{w}_{i}\!\times\!\mathbf{w}_{j})\!\cdot\!(\mathbf{w}_{k}\!\times\!\mathbf{w}_{l}) =(\mathbf{w}_{i}\!\cdot\!\mathbf{w}_{k})(\mathbf{w}_{j}\!\cdot\!\mathbf{w}_{l})-(\mathbf{w}_{i}\!\cdot\!\mathbf{w}_{l})(\mathbf{w}_{j}\!\cdot\!\mathbf{w}_{k}) \;,
\ee
which we separate into $\mathbb{P}_{\rm dir}^{22}=\mathbb{P}_{\rm dir}^{22\to2}+\mathbb{P}_{\rm dir}^{22\to1}$, where $\mathbb{P}_{\rm dir}^{22\to2}$ and $\mathbb{P}_{\rm dir}^{22\to1}$ are obtained by replacing $\theta(\theta_{31})\theta(\theta_{42})$ with the first and second term in
\be\label{StepsForStepsSep}
\theta(\theta_{42})\theta(\theta_{31}) 
=\theta(\sigma_{43}-\sigma_{21})\left\{1-\theta\left(\frac{|\theta_{43}-\theta_{21}|}{2}-[\sigma_{43}-\sigma_{21}]\right)\right\} \;, 
\ee 
respectively,
and
\be
\label{P22exGen}
\begin{split}
	\mathbb{P}_{\rm ex}^{22}(s)=\text{Re}\frac{\alpha^2}{16\pi^2b_0^2}&\int\!\ud\phi_{1234}\frac{\theta(\theta_{42})\theta(\theta_{31})}{s_0s_1s_2s_3d_0} 
	\Bigg\{F_0+f_0-\frac{2ib_0}{d_0}(f_1+z_1)+\left[\frac{2b_0}{d_0}\right]^2z_2\Bigg\} \\
	&\exp\left\{\frac{i}{2b_0}\frac{q_1q_2}{s_0s_1s_2s_3d_0}\left(\theta_{41}\theta_{23}\left[\frac{\Theta_{41}}{s_1}+\frac{\Theta_{23}}{s_2}\right]+\theta_{43}\theta_{21}\left[\frac{\Theta_{43}}{s_3}-\frac{\Theta_{21}}{s_0}\right]+\theta_{31}\theta_{42}\left[\frac{\Theta_{31}}{q_2}-\frac{\Theta_{42}}{q_1}\right]\right)\right\} \;.
\end{split}
\ee
The prefactor of~\eqref{P22exGen} can be expressed in terms of a permutation defined by
\be\label{PermutationDefinition}
{\sf P}[\mathcal{F}]:=\mathcal{F}(\phi_1\to\phi_2\to\phi_3\to\phi_4\to\phi_1,s_1\to-s_0\to s_2\to s_3\to s_1) \;,
\ee 
as
\be
{\bf d}_1=\frac{1}{d_0}\left(\frac{\theta_{23}}{s_2}\frac{\theta_{41}}{s_1}{\bf\Delta}_{14}+\frac{\theta_{21}}{s_0}\frac{\theta_{43}}{s_3}{\bf\Delta}_{12}+\frac{\theta_{42}\theta_{23}}{s_2s_3}[{\bf\Delta}_{24}-{\bf\Delta}_{23}]\right) 
\qquad
d_0=\frac{\theta_{23}}{s_2}\frac{\theta_{41}}{s_1}+\frac{\theta_{21}}{s_0}\frac{\theta_{43}}{s_3} \;,
\ee
\be
{\bf d}_2={\sf P}[{\bf d}_1] \qquad {\bf d}_3={\sf P}[{\bf d}_2] \qquad {\bf d}_4={\sf P}[{\bf d}_3] \qquad {\bf d}_1={\sf P}[{\bf d}_4] \;.
\ee
\be
F_0
=(1+{\sf P})\kappa_{03}[({\bf d}_1\!\cdot\!{\bf d}_3)({\bf d}_2\!\cdot\!{\bf d}_4)+(\mathbf{d}_{1}\!\times\! \mathbf{d}_{3})\!\cdot\!(\mathbf{d}_{2}\!\times\! \mathbf{d}_{4})] \;,
\ee
\be
\begin{split}
	f_0&=(1+{\sf P})\frac{1}{s_0s_1s_2s_3}(s_1q_2{\bf d}_1-s_2q_1{\bf d}_2)\!\cdot\!(s_2q_2{\bf d}_3-s_1q_1{\bf d}_4)\;, \\
	f_1&=-(1+{\sf P}+{\sf P}^2+{\sf P}^3)\kappa_{03}\frac{\theta_{21}}{s_0}{\bf d}_2\!\cdot\!{\bf d}_1+(1+{\sf P})(\kappa_{03}-\kappa_{12})\frac{\theta_{42}}{q_1}{\bf d}_4\!\cdot\!{\bf d}_2 \;,
\end{split}
\ee 
\be
z_1
=(1+{\sf P}+{\sf P}^2+{\sf P}^3)\frac{-q_1^2}{s_0s_1q_2}\left(3+\frac{s_2s_3}{s_0s_1}\right)\phi_1 
\qquad
z_2
=(1+{\sf P})\kappa_{03}\left(\frac{\theta_{43}}{s_3}\frac{\theta_{21}}{s_0}+\frac{\theta_{31}}{q_2}\frac{\theta_{42}}{q_1}\right)\;.
\ee

\end{widetext}

We have obtained these results by performing the Gaussian integrals over $p_{1\LCperp}$ and $p_{2\LCperp}$. For the direct terms we find that the Gaussian peak is located at
\be
p_1^\LCperp=\langle a\rangle_{21}^\LCperp+s_1(p^\LCperp-\langle a\rangle_{21}^\LCperp)
\ee 
and
\be
p_2^\LCperp=\langle a\rangle_{43}^\LCperp+s_2(p^\LCperp-\langle a\rangle_{21}^\LCperp) \;,
\ee 
which means
\be
p_3^\LCperp=-\langle a\rangle_{43}^\LCperp+s_3(p^\LCperp-\langle a\rangle_{21}^\LCperp) \;.
\ee
It is experimentally relevant to assume that $\gamma=p_0$ is much larger than all the other parameters, and then we have to leading order 
\be\label{parallelparticles}
{\bf p}_i\approx s_i{\bf p} \;.
\ee 
The peak for the exchange terms is in general more complicated, but reduces~\eqref{parallelparticles} to leading order.
Thus, in this regime all particles in the final state have momenta approximately parallel to the initial momentum.
$s_i$ is by definition the fraction of the initial longitudinal momentum given to particle $i$. From~\eqref{parallelparticles} we see that it also gives the fraction of the total momentum to leading order. $s_i$ are therefore natural variables to focus on. After shifting and diagonalizing the transverse momentum variables, the exponential becomes $e^{-c_1 p_{1\LCperp}^2-c_2 p_{2\LCperp}^2}$. The coefficients $c_1$ and $c_2$ determine the width of the Gaussian peak. $c_i$ can be expressed in terms of $s_i$ and $\phi_i$ and scale with $a_0$ and $b_0$, but do depend on $\gamma$ separately ($b_0$ is much smaller than $\gamma$). The width of the Gaussian peak is therefore much smaller than its position.

\end{document}